\documentclass[11pt]{iopart}
\usepackage[dvips]{graphicx}
\usepackage{here}  
\begin{document}

\title[Charmonia enhancement in quark-gluon plasma]{Charmonia enhancement in 
quark-gluon plasma with improved description of c-quarks phase-distribution}

\author{Pol Bernard Gossiaux\dag
\footnote[3]{To whom correspondence should be addressed 
(pol.gossiaux@subatech.in2p3.fr)},
Vincent Guiho\dag, and~J\"org Aichelin\dag}

\address{\dag\ SUBATECH, \'Ecole des Mines de Nantes, 4 rue Alfred Kastler, 
44307 Nantes Cedex 3, France}

\begin{abstract}
We present a dynamical model of heavy quark evolution in the quark-gluon 
plasma (QGP) based on the Fokker-Planck equation. We then apply this model to 
the case of central ultra-relativistic nucleus-nucleus collisions performed at
RHIC and estimate the component of $J/\psi$ production (integrated and 
differential) stemming from c-$\bar{c}$ pairs that are initially uncorrelated. 
\end{abstract}




\section{Introduction}
According to the first experimental results on open charm production in 
nucleon-nucleon reactions at RHIC $(\sqrt{s}=200~{\rm GeV})$, up to 40 
c-$\bar{\rm c}$ pairs could be formed in a central Au-Au collision at 
this energy. As noticed in \cite{thews}, this large number could 
result in the formation of additional charmonia as compared to a simple
scaling of nucleon-nucleon prompt production\footnote{For a more detailed 
discussion of these ideas, see the contribution of R.L. Thews to this 
conference.}, due to ${\rm c}+\bar{\rm c} \rightarrow \psi + X$ 
reactions happening later in the QGP. One would then observe a charmonia 
{\em enhancement} instead of suppression.  However (i) 
${\rm D}+\bar{\rm D}\rightarrow \psi +{\rm X}$ reactions happening during the 
ensuing hadronic phase can also contribute to charm enhancement and (ii) such 
an enhancement does not seem to be observed experimentally 
(preliminary $J/\psi$ data appear to be compatible with the usual suppression).
All of this indicates that the models should be refined.
In fact, the actual distribution of ${\rm c}$ and $\bar{\rm c}$ in 
phase-space could have a large impact on these recombination processes, but is
often not treated  with the care it deserves (or is simply beyond the scope of
some models). Here, we will complement the model introduced in \cite{thews} by
a dynamical treatment of the heavy-quarks evolution in the QGP based on a 
Fokker-Planck equation. We also present and discuss the differential spectra 
of c quarks and charmonia at the end of the QGP phase.

\section{The model}
Following \cite{svet,must}, the $\rm c$ and $\bar{\rm c}$ quarks 
distributions in the 
QGP are assumed to follow a Fokker-Planck (FP) equation in momentum space.
\begin{equation}
\frac{\partial f(\vec{p},t)}{\partial t}=\frac{\partial}{\partial p_i}
\left[A_i(\vec{p}) f(\vec{p},t)+
\frac{\partial }{\partial p_j}\left(B_{ij}(\vec{p}) f(\vec{p},t)
\right)\right]
\end{equation} 
The main justification for this hypothesis is that the heavy mass of these 
quarks implies large relaxation times as compared to the typical time of 
individual ${\rm c}+{\rm g}\rightarrow {\rm c}'+{\rm g}'$ and ${\rm c}+
{\rm q}\rightarrow {\rm c}'+{\rm q}'$ collisions, whatever the momentum of the
heavy quark. The drag ($A$) and diffusion ($B$) coefficients were evaluated 
according to
\cite{svet,must,cley}, resorting to a power expansion 
``\`a la Kramers - Moyal'' (KM) of the Boltzmann kernel of $2\rightarrow 2$ 
collisions.  As realized in \cite{rafel}, the asymptotic distribution coming
out of the FP evolution deviates from a Boltzmannian. This results from the 
truncation of the KM series. We
have therefore slightly corrected the $B$ coefficients in order to guarantee
a correct Boltzmannian asymptotic distribution. We have also checked on
various examples that this modification has little effect at small
evolution time. This constitutes our reference set of FP coefficients. In 
order to circumvent our lack of knowledge on the c-QGP interaction (radiative 
processes, non perturbative aspects, etc.\footnote{Ultimately, transport 
coefficients should be evaluated using lattice calculations.}), we will also
consider alternative sets, obtained by multiplying the reference one by a
numerical factor $K$. It is our hope that the gross experimental results will 
permit to fix an approximate value of $K$ and that our model could then be 
used to predict finer aspects. In this respect, we view our model as a 
semi-predictive effective theory that could be useful to match the gap 
between the fundamental underlying theory (QCD) and experimental results.

The FP coefficients depend on position and time (only) through the local 
temperature and velocity of the surrounding medium, that is assumed to be 
locally thermalized and described via hydrodynamic evolution. Therefore,
we do need to evaluate explicitly all microscopic 
${\rm c}+{\rm g}\rightarrow {\rm c}'+{\rm g}'$ and 
${\rm c}+{\rm q}\rightarrow {\rm c}'+{\rm q}'$ processes when performing our 
simulations: only ${\rm c}$, $\bar{\rm c}$ and $J/\psi$ d.o.f. are considered.

For the sake of simplicity, we have neglected all initial state interactions
in this exploratory work. We have also allowed ourselves to take a simple 
factorized form of the initial ${\rm c}-\bar{\rm c}$ phase-distribution, 
{\it i.e.} $f_{\rm in}(\vec{r}_{\rm c},\vec{p}_{\rm c};\vec{r}_{\bar{\rm c}},
\vec{p}_{\bar{\rm c}})\propto
T_{A}(\vec{r}_{{\rm c},\perp}) T_{B}(\vec{r}_{\bar{\rm c},\perp})
\delta^{(3)}(\vec{r}_{\rm c}-\vec{r}_{\bar{\rm c}})
\delta(z_{\rm c})\times f_{\rm in}(\vec{p}_{\rm c})\times 
f_{\rm in}(\vec{p}_{\bar{\rm c}})$
with $f_{\rm in}(\vec{p}_{\rm c})\propto 
f_{\rm in}(\vec{p}_{{\rm c},\perp})\times
f_{\rm in}(y_{c})$,
where $f_{\rm in}(y_{\rm c})$ has been chosen according to \cite{vogt} and 
$f_{\rm in}(\vec{p}_{{\rm c},\perp})$ according to $\rm D$ transverse-momentum
spectra at mid-rapidity in nucleon-nucleon reactions. 

$J/\psi$ mesons will be assumed to exist up to a dissociation temperature
$T_{\rm diss}$ that we let as a free parameter. Under this temperature, they 
behave as if they were in vacuum. According to recent lattice calculation, 
$T_{\rm diss}$ could be substantially above the transition temperature.   
Apart from the prompt component, $J/\psi$ are assumed to be formed and 
destroyed in the QGP via the ${\rm c}+\bar{\rm c}\leftrightarrow J/\psi + 
{\rm g}$ processes \cite{thews}. In our simulation, these radiative 
dissociations are implemented via local dissociation rates that efficiently 
incorporate the microscopic processes. 

\section{Results and discussion}
All results are for central Au+Au reactions at $\sqrt{s}=200~{\rm GeV}$.
The initial number $N_{\rm c}$ of ${\rm c}$ and $\bar{\rm c}$ quarks is taken 
to be 40, with $\frac{{\rm d}N_{\rm c}(y=0)}{{\rm d}y}\approx 9$. Unless 
specified differently, the temperature and velocity profiles have been 
evaluated with the hydrodynamic model of Kolb and Heinz \cite{kolb}. 
In order to address most clearly the effect of the uncorrelated $J/\psi$ 
production, the prompt $J/\psi$ component has been set to zero.

\begin{figure}[H]
\begin{center}
\includegraphics[height=4.cm] {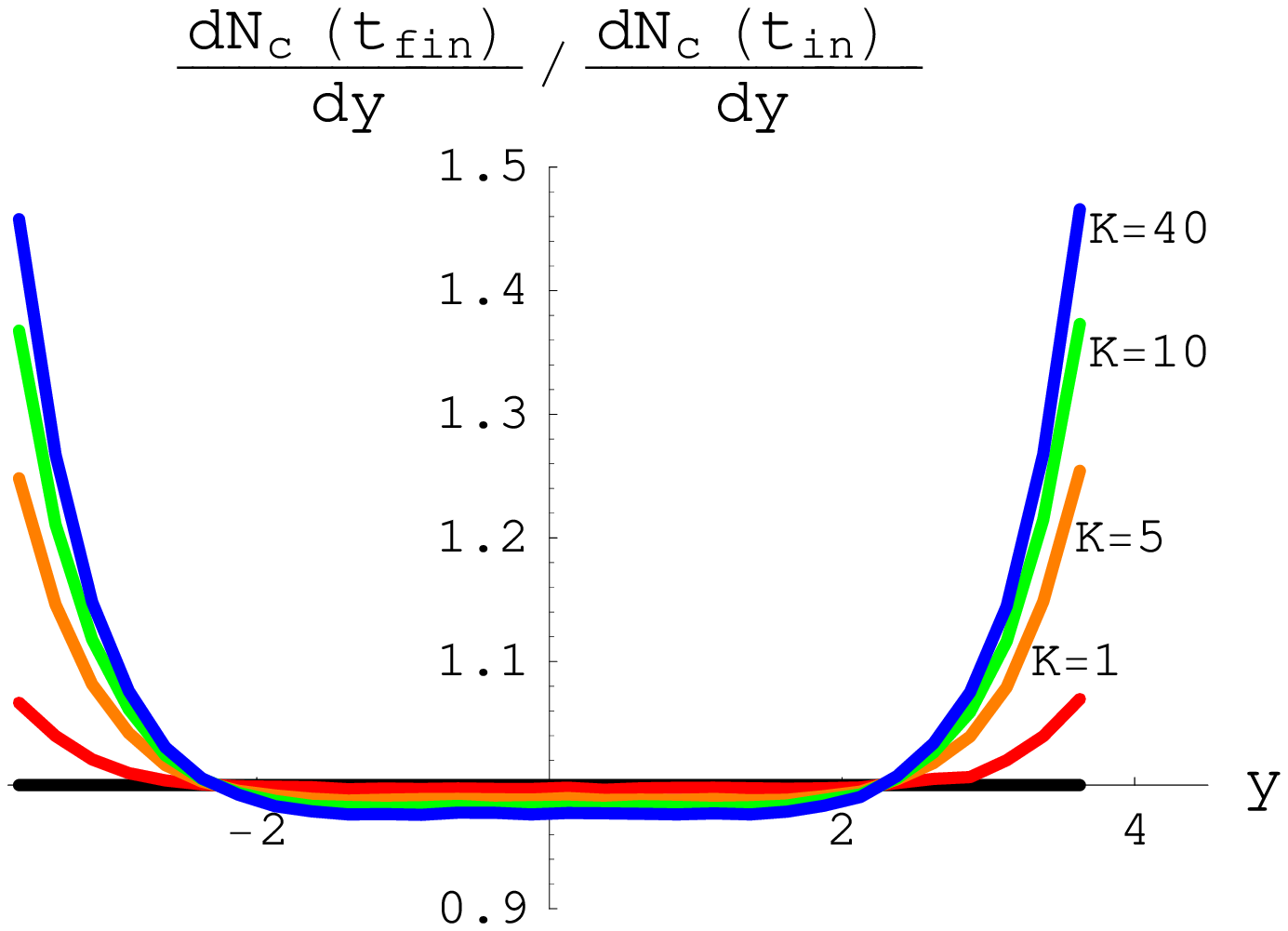} 
\includegraphics[height=4.cm] {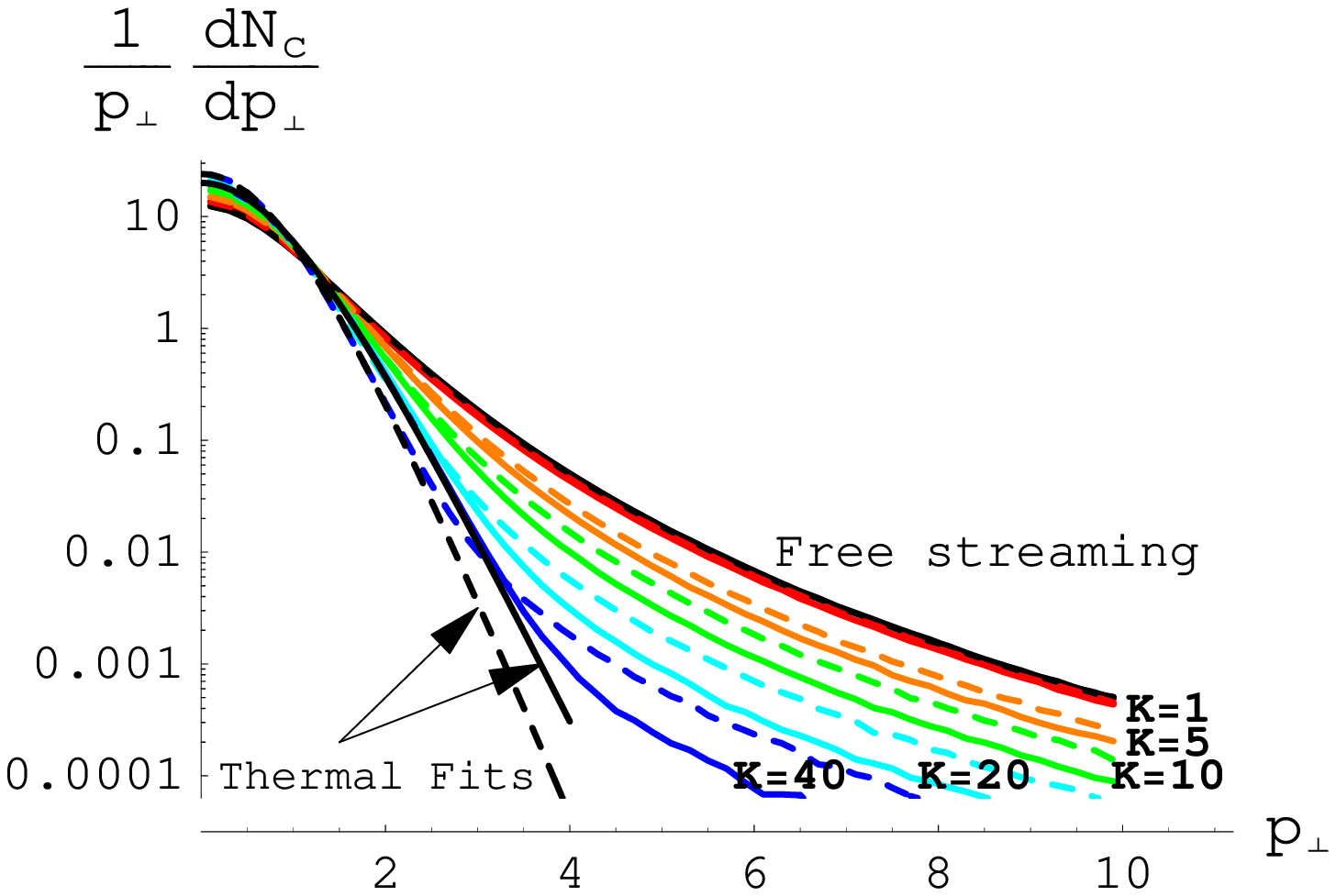} 
\end{center}
\caption{\label{fig1} differential productions of c quarks at the end
of the QGP phase: On the left panel, the rapidity distribution is 
presented relative to the one at initial time. On the right panel, transverse
momentum spectra are displayed for Kolb and Heinz hydrodynamic evolution
of QGP (full lines), as well as for some Bjorken hydrodynamic evolution 
with no transverse expansion (broken lines).}
\end{figure}

We show the differential productions profile of $c$ quarks at the end
of the QGP phase in figure \ref{fig1} for different values of $K$.
Let us remind that typical relaxation times are $\propto K^{-1}$. Thus, $K=0$ 
corresponds to the free streaming case while $K\rightarrow +\infty$ 
corresponds to instantaneous equilibration of heavy quarks with the QGP.   

On a broad window around mid-rapidity, the final profile deviates merely from 
the initial one. One just notices a small broadening for large values of $K$. 
Indeed, heavy quarks locate themselves very early in a slot of matter going at
the same rapidity. The drag force thus vanishes and the heavy quarks conserve 
their rapidity for the rest of the evolution. However, diffusion processes can
still happen and bring heavy quarks towards the tail of the rapidity 
distribution, leading to some rapidity broadening.

As for the transverse momentum distributions, the noticeable fact is
the ``cooling'' of c quarks transverse motion by the QGP. One observes 
significant deviations from the free streaming case for $K\ge 5$ only. For 
large values of $K$, distributions tend to a Boltzmannian at small and 
intermediate values of $p_\perp$. As the distributions associated to different
values of $K$ interpolate smoothly between those two limiting cases,
it might be possible -- once initial and final state effects are included -- 
to ``read'' the value of $K$ from experimental spectra of $D$ mesons. 

\begin{figure}[H]
\begin{center}
\includegraphics[height=4.cm] {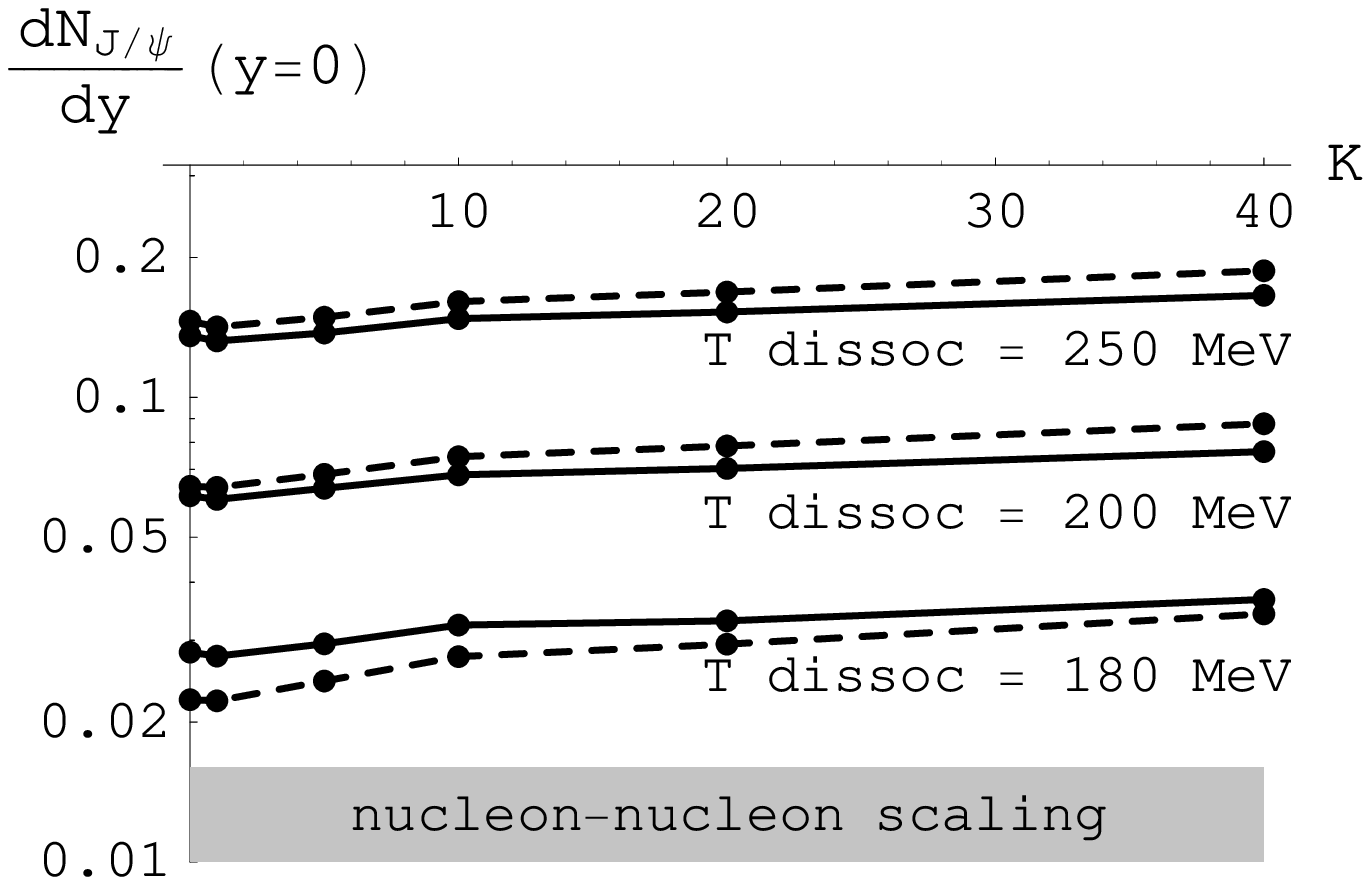} 
\includegraphics[height=4.cm] {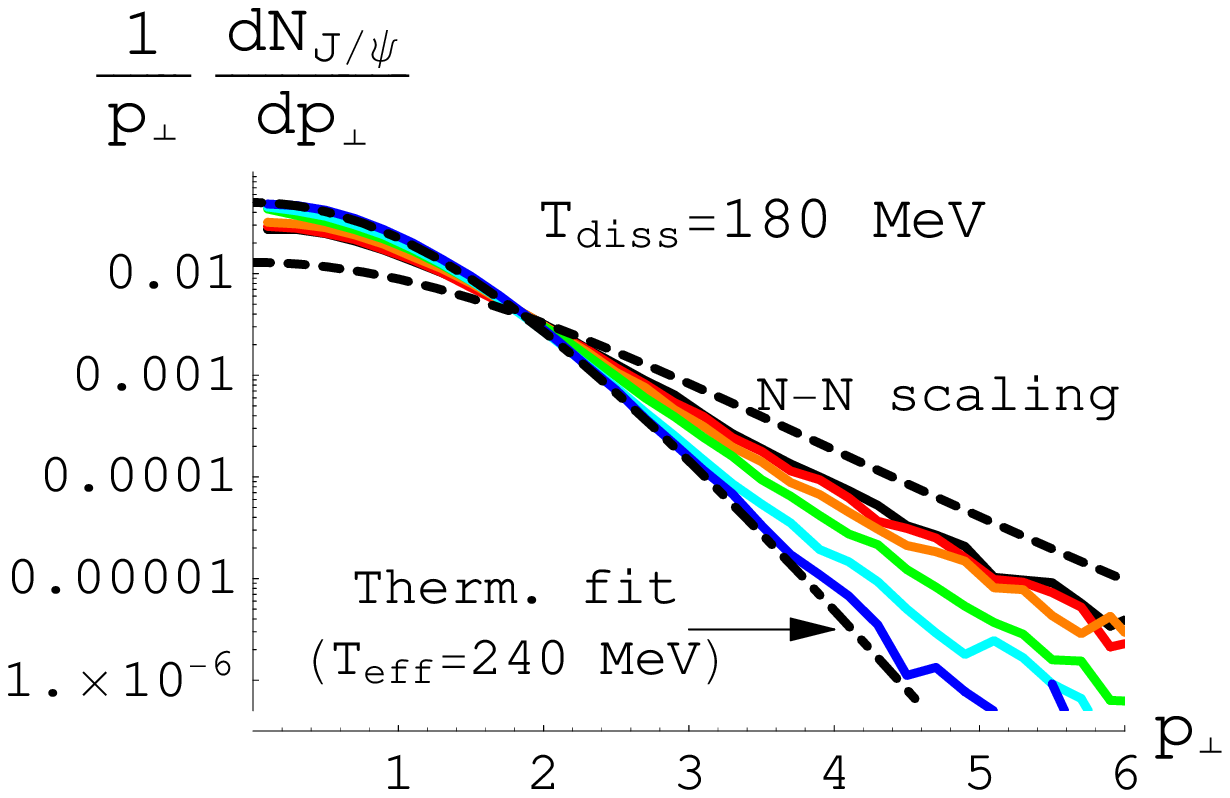} 
\end{center}
\caption{\label{fig2} Left: $J/\psi$ production at mid-rapidity at the 
end of the QGP phase with (full) and without (broken) radial 
expansion. Right: transverse momentum spectrum at mid-rapidity;
plain lines correspond to $K=0$ (hardest spectrum at large $p_\perp$), 1, 5, 
10, 20 \& 40. The chain curve represents the best fit of the $K=40$ 
spectrum by a Boltzmannian, while the broken curve represents the prompt 
$J/\psi$ component.}
\end{figure}

We present the differential production of $J/\psi$ at mid-rapidity on the left
panel of figure \ref{fig2}. $T_{\rm diss}$ appears to have a much larger 
influence on this quantity than $K$. Once the initial number of 
${\rm c}-\bar{\rm c}$ is fixed more accurately, it might be possible to assess
the value of $T_{\rm diss}$ and perform comparison with theoretical estimates.
The right panel shows some typical predictions for $J/\psi$ transverse 
spectrum at mid-rapidity. In all cases -- but especially for large $K$ -- the 
prompt component is harder that the uncorrelated one.

\section{Conclusion and outlook}
We have presented a model that copes efficiently with dynamical 
evolution of heavy quarks and quarkonia in QGP. We have evaluated
$J/\psi$ production at mid-rapidity as well as some differential spectra  
for different equilibration ``strengths'' $K$ and for different dissociation
temperatures. Hopefully, the comparison of such results with their 
experimental equivalent will ultimately  allow to ``measure'' these 
parameters characteristic of the QGP phase.

\section*{Acknowledgments}
The authors are most grateful to E. Frodermann, P. Huovinen, M.G. Mustafa, and 
D. Rischke for having provided them with their respective codes or valuable 
data.

\end{document}